\documentclass[a4paper,11pt]{article}
\pdfoutput=1 % if your are submitting a pdflatex (i.e. if you have
\usepackage{jcappub} % for details on the use of the package, please
\usepackage[utf8]{inputenc}
\usepackage{graphicx}
\usepackage{subcaption}
\def\degree{$^{\circ}$}
\usepackage{lineno}

\title{Search for the evaporation of primordial black holes with H.E.S.S.}

\author[1,2]{F.~Aharonian,}
\author[3]{F.~Ait~Benkhali,}
\author[4]{J.~Aschersleben,}
\author[5]{M.~B\"ottcher,}
\author[6,5]{M.~Backes,}
\author[7]{V.~Barbosa~Martins,}
\author[8]{R.~Batzofin,}
\author[9,10]{Y.~Becherini,}
\author[7,11]{D.~Berge,}
\author[12]{B.~Bi,}
% ...

\affiliation[1]{Dublin Institute for Advanced Studies, 31 Fitzwilliam Place, Dublin 2, Ireland}
\affiliation[2]{Max-Planck-Institut f\"ur Kernphysik, P.O. Box 103980, D 69029 Heidelberg, Germany}
\affiliation[3]{Landessternwarte, Universit\"at Heidelberg, K\"onigstuhl, D 69117 Heidelberg, Germany}
\affiliation[4]{Kapteyn Astronomical Institute, University of Groningen, Landleven 12, 9747 AD Groningen, The Netherlands}
\affiliation[5]{Centre for Space Research, North-West University, Potchefstroom 2520, South Africa}
\affiliation[6]{University of Namibia, Department of Physics, Private Bag 13301, Windhoek 10005, Namibia}
\affiliation[7]{DESY, D-15738 Zeuthen, Germany}
\affiliation[8]{School of Physics, University of the Witwatersrand, 1 Jan Smuts Avenue, Braamfontein, Johannesburg, 2050 South Africa}
\affiliation[9]{Université de Paris, CNRS, Astroparticule et Cosmologie, F-75013 Paris, France}
\affiliation[10]{Department of Physics and Electrical Engineering, Linnaeus University,  351 95 V\"axj\"o, Sweden}
\affiliation[11]{Institut f\"ur Physik, Humboldt-Universit\"at zu Berlin, Newtonstr. 15, D 12489 Berlin, Germany}
\affiliation[12]{Institut f\"ur Astronomie und Astrophysik, Universit\"at T\"ubingen, Sand 1, D 72076 T\"ubingen, Germany}
% ...

\abstract
{
Primordial Black Holes (PBHs) are hypothetical black holes predicted to have been formed from density fluctuations in the early Universe. 
PBHs with an initial mass around $10^{14}-10^{15}$g are expected to end their evaporation at present times in a burst of particles and very-high-energy (VHE) gamma rays. 
Those gamma rays may be detectable by the High Energy Stereoscopic System (H.E.S.S.), an array of imaging atmospheric Cherenkov telescopes. 
This paper reports on the search for evaporation bursts of VHE gamma rays with H.E.S.S., ranging from 10 to 120 seconds, as expected from the final stage of PBH evaporation and using a total of 4816 hours of observations.
The most constraining upper limit on the burst rate of local PBHs is 
$2000$ pc$^{-3}$ yr$^{-1}$ for a burst interval of 120 seconds, at the 95\% confidence level. The implication of these measurements for PBH dark matter are also discussed.
}

\date{\today}

\begin{document}
\maketitle
%\linenumbers

\section{Introduction}\label{sec:intro}
Primordial black holes (PBHs) \cite{1974MNRAS.168..399C} have been predicted to form in the early Universe via a variety of mechanisms \cite{2005astro.ph.11743C}.  
The gravitational collapse of over-dense regions with significant density fluctuations \cite{1997PhRvD..56.6166G} is the best known of these
mechanisms. It requires a spectrum of primordial density fluctuations with an excess of power on small spatial scales to be efficient. Other formation mechanisms include pressure reduction during cosmological phase transitions \cite{1997PhRvD..55.5871J}. The mass function of PBHs depends on the actual formation mechanism. PBHs could have masses ranging from 10$^{-5}$g for PBHs created at the Planck time up to several tens of $M_{\odot}$ for PBHs created during the QCD phase transition \cite{1997PhRvD..55.5871J}.    
No PBH candidate has been unambiguously detected, 
%although there has been claims of such detections, notably progenitors of recent black hole merging events \cite{2016PhRvL.116t1301B} and MACHOs \cite{2016PhRvD..94f3530G}. 
although recent black hole merging events \cite{2016PhRvL.116t1301B}  and Massive Compact Halo Objects (MACHOs) \cite{1997PhRvD..55.5871J}
could be explained with PBHs.

Black holes 
were predicted by Hawking \cite{1974Natur.248...30H} to radiate particles with a black body spectrum 
%of energies. The emission can be described by an
of effective temperature
\begin{equation}
 T_{\mathrm{BH}} = \frac{M_p^{2}}{8\pi M_{\mathrm{BH}}}\,,
\label{eq:Tvsmass}
\end{equation}
where $\mathrm{M_p}$ and $M_{\mathrm{BH}}$ are the Planck mass and the PBH mass, respectively. In equation (\ref{eq:Tvsmass}) and in the following, the temperature and the masses are written in energy units, the constants $k_{B}, \hbar$ and $c$ are all set to 1.
%Equation (\ref{eq:Tvsmass}) has a simple quantum mechanical interpretation: radiation of 
%wavelength $\lambda = 2\pi/ T_{\mathrm{BH}},$ typical of a black-body with temperature $T_{\mathrm{BH}}$ (Wien's law), cannot be localized inside a black hole of size $R_{\mathrm{BH}} = 2G M_{\mathrm{BH}}$ if $R_{\mathrm{BH}} \le \lambda.$  This
%can be seen as a consequence of Heisenberg inequalities \cite{2014arXiv1402.1427C}.
Hawking's radiation is 
%therefore 
negligible for black holes of stellar masses or higher. However evaporation becomes the predominant process that governs the black hole evolution for small-mass black holes. Black holes lose their mass by Hawking radiation at a rate inversely proportional to their squared mass.

A popular method for constraining the density of low-mass PBHs is searching for their gamma-ray emission. Searches have attempted to detect a diffuse photon signal from a distribution of PBHs \cite{2016PhRvD..94d4029C} or to look directly for the final stage emission of an individual hole \cite{2017ICRC...35..691A,2018ApJ...857...49A, 2020JCAP...04..026A}.

The present paper reports on the search for TeV gamma-ray bursts with a timescale of a few seconds to a few minutes\cite{1991Natur.353..807H}, as expected from the final stage of PBHs evaporation, using the H.E.S.S. array of Imaging Atmospheric Cherenkov Telescopes (IACTs). % (Sec. \ref{sec:hess}). 
The H.E.S.S. data selection and processing are presented in Sec. \ref{sec:dataset}.
The modelling of the PBH signal is described in Sec. \ref{sec:model}. The burst search algorithm and background estimation are discussed in Sec. \ref{sec:analysis}. The results on the local evaporation rate, and the comparison to existing limits are given in Sec. \ref{sec:results}. The implication of the obtained results on PBH dark matter is discussed in Sec. \ref{sec:discussion}. Finally, conclusions are drawn in Sec. \ref{sec:conclusion}.

\section{H.E.S.S. array and data set}\label{sec:dataset}%\label{sec:hess}
The High Energy Stereoscopic System (H.E.S.S.) is an array of five imaging atmospheric Cherenkov telescopes dedicated to observing very-high-energy (VHE; $\gtrsim$~100~GeV) gamma rays from astrophysical sources. 
It is located in the Khomas Highland of Namibia at an altitude of 1800 m above sea level. 
The first four telescopes were installed in 2001-2003 (H.E.S.S.1 phase of the experiment) and have been operational since 2004. Each telescope of H.E.S.S.1 is equipped with a tessellated optical reflector of $107\  \mathrm{m}^{2}$ \cite{2003APh....20..111B} and a camera with 960 photomultiplier tubes. The camera field of view is $5^{\circ}$ in diameter. 
The stereoscopic technique %\cite{2004APh....22..285F} 
\cite{1997APh.....8....1D}
allows for an
accurate reconstruction of the direction and the energy of the primary 
gamma ray. H.E.S.S.1 has an angular resolution of less than $0.1$\degree, 
a source location accuracy of   $\sim 30''$ for strong sources and an effective detection area of $\sim 10^{5}\ \mathrm{m}^2$.
The sensitivity for point-like sources reaches $2\times 10^{-13}\ 
\mathrm{cm}^{-2}\mathrm{s}^{-1}$ above 1 TeV for a 5$\sigma$ detection in 25 hours of a source at
a $20^{\circ}$ zenith angle \cite{2009APh....32..231D}.
The fifth telescope with a reflective area of $614\ \mathrm{m}^2$, which started its operations in 2012, is not used in this analysis.  

%\section{Dataset}\label{sec:dataset}

%% VIM : Rephrase a bit the sentence below ?
 The data used for this analysis are the vast majority of the H.E.S.S. observations taken between January 2004 and January 2013. 
Each observation is called a run and consists of data taken pointing towards the same position in the sky during $\sim 28$ minutes. 
The runs with poor quality (due to hardware problems, bad atmosphere, etc...) as well as those with large zenith angle ($>60^{\circ}$) were removed in order to reduce the systematic uncertainties. This yields a total of 11234 runs which corresponds to 4816 hours.
%After removing poor quality runs (due to hardware problem or bad atmosphere), a total of 11234 runs were selected, corresponding to 4816 hours.
%VIM 15/11/2022 above is the rephreasing of this paragraph (ends with %%%%% END REPHRASING)
% The data used for this analysis are all the H.E.S.S. observations taken between January 2004 and January 2013. %towards sky position with Galactic latitudes $|b| > 10^{\circ}$. 
% One H.E.S.S. observation run consists of data taken pointing towards the same position on the sky during $\sim 28$ minutes. 
% Some regions of the sky (Crab, Large and Small Magellanic Clouds, region of SN 1006) were excluded from this analysis as were runs of poor quality, affected for instance by bad weather or technical problems. 
% The data set
% consists of 
% %11494 %Before VIM Analysis
% 11234 %VIM Analysis
% runs, corresponding to 
% %more than
% %4924 % Before VIM Analysis
% 4816 % VIM Analysis
% hours of observations. 
%The data have been processed by two independent calibration and reconstruction chains.
%%%%% END REPHRASING
%% Another rephrasing possibility:
% VIM : Original line : The ImPACT analysis \cite{2014APh....56...26P} was applied to all the runs in order to suppress the background of hadronic cosmic rays and reconstruct the direction and energy of the gamma-ray candidates. 
The event reconstruction (direction, energy) has been perform using the ImPACT method \cite{2014APh....56...26P} and the identification of background events was done using boosted decision trees \cite{2009APh....31..383O}.
%The arrival times of these events so-called ``gamma-like'' events are extracted together with their reconstructed parameters.  %% VIM 15/11/2022 : Remove this sentence, not really useful.
%For each run, gamma-like 
Events with a distance to the center of the camera larger than $2$\degree are excluded. In addition, events for which the energy bias is estimated to be above $10\%$ are excluded, defining a safe energy range for each run.
The results were cross-checked 
%with the Model analysis 
with a different calibration and event reconstruction pipelines
\cite{2009APh....32..231D}, giving consistent results.

\section{Expected PBH signal}\label{sec:model}
To discover or constrain PBH evaporation, observations have to be compared to models.
During the evaporation process, the PBH heats up by emitting particles with masses around and below its temperature.
The  prediction of the instantaneous gamma-ray spectrum d$^{2}$N/dE$_{\mathrm{\gamma}}$dt 
emitted by the PBH in its last seconds  depends on the assumed elementary particle 
mass spectrum and more generally on the extrapolation of known physics to large PBH temperatures
\cite{1991Natur.353..807H}.
In the hadronic model of Hagedorn \cite{1968NCimA..56.1027H}, temperature is limited by an ultimate value of $\simeq 160$ MeV. In this model, the signature of PBH evaporation is a microsecond burst of photons with energies $\simeq 100$ MeV \cite{2009APh....31..102S}.
In loop quantum gravity, the PBH might explode before total evaporation, leading to short $\sim 10\ \mathrm{MeV}$ bursts \cite{2014PhLB..739..405B}. 
%In other models, such as the standard model of %particle physics, PBH temperatures
In the more standard MacGibbon-Webber model \cite{1990PhRvD..41.3052M} , the black hole emits only those Standard Model species which appear fundamental on the scale of the black hole temperature, decaying and hadronizing as they stream away from the black hole, and so PBH temperatures in the TeV range can be reached. In supersymmetric particle models, additional degrees of freedom tend to accelerate the PBH evaporation \cite{2016APh....80...90U}. 
The gamma rays detectable by H.E.S.S. are emitted when the PBH has a temperature in the 100 GeV-10 TeV energy range. 
It is assumed in this paper that the standard model of particle  physics remains valid in the H.E.S.S. temperature range.  
Another systematic effect in the modelling is the presence of a photosphere around the PBH \cite{1997PhRvD..55..480H,2002PhRvD..65f4028D}, which would drastically alter the 
evaporation signal 
by suppressing the high-energy component of emitted particles. However, photospheric effects around PBHs are under debate \cite{2008PhRvD..78f4043M}. In this paper, 
possible PBH photosphere effects on the evaporation signal have been neglected.

Following the analysis in \cite{1991Natur.353..807H} and \cite{2006JCAP...01..013L}, the integrated 
spectrum of photons per square meter emitted during the time $\Delta$t before total evaporation of the PBH
and observed by a detector at a distance $r_0$ 
%above the energy $E$ 
is given by :
\begin{equation}
N(>E)=
%2.4\ 10^{37} 
0.22 \left(\frac{0.1\ \mathrm{pc}}{r_0}\right)^2
\left(\frac{\mathrm{TeV}}{Q}\right)^{2}\left(\frac{5}{14}\left(\frac{E}{Q}\right)^{\frac{3}{2}}+3\left(\frac{E}{Q}\right)^{1/2}+\frac{5}{6}\left(\frac{Q}{E}\right)^{1/2}-\frac{5}{3}\left(\frac{E}{Q}\right)-\frac{5}{2}+\frac{1}{150}\right)\,,
\label{eq:lowE}
\end{equation}
for $E < Q$ and by :
\begin{equation}
N(>E)=
0.22 \left(\frac{0.1\ \mathrm{pc}}{r_0}\right)^2
%2.4\ 10^{37}
\left(\frac{\mbox{TeV}}{E}\right)^{2}\left(\frac{1}{42}+\frac{1}{150}\right)\,,
\label{eq:highE}
\end{equation}
for $E \ge Q.$
In the above equations, $E$ is the energy, and $Q$  relates to $\Delta t$ by: 
\begin{equation}
Q\simeq 40\ \mbox{TeV} (1\ \mbox{s}/\Delta t)^{1/3}.
\label{eq:Qdef}
\end{equation}
$Q$ is a slowly decreasing function of $\Delta t,$ with values for the observations reported in this paper ranging from $\sim 8$ TeV for $\Delta t =120$ seconds to  $\sim 13$ TeV for $\Delta t =30$ seconds.  H.E.S.S. is sensitive to PBH evaporations up to distances of order $r_0 = 0.1$ pc, so that the expected flux does not need to be corrected for Extragalactic Background Light (EBL) absorption.\\  
Instead of equation \ref{eq:lowE}, the low-energy asymptotic form
\begin{equation}
N(>E)=
%2.4\ 10^{37} 
0.18 \left(\frac{0.1 \mathrm{pc}}{r_0}\right)^2
\left(\frac {\mbox {TeV}}{Q}\right)^{2}\left(\frac{Q}{E}\right)^{1/2}
\label{eq:lowEsimple}
\end{equation}
was used in the predictions for convenience.   
This simple power-law form is very close to the fit of the MacGibbon-Weber model \cite{1990PhRvD..41.3052M} given in \cite{2008ICRC....3.1123B}. 
Equations \ref{eq:lowE}-\ref{eq:lowEsimple} can 
be simply understood as describing a broken power-law spectrum with a low-energy photon index of $\frac{5}{2},$ a high-energy photon index of 3 and a break energy of $Q$.  

% Comparison with other calculations
%Our spectrum has also been compared to more recent calculations in \cite{2016APh....80...90U} and to the output of
%the public BlackHawk program \cite{2019EPJC...79..693A}. The spectra are consistent at the factor 2 level in the H.E.S.S. energy range. 
% VIM : 15/11/2022 : Rephrase attempt : 
Our spectrum, when compared to more recent analytical calculations in \cite{2016APh....80...90U} and to the output of the public BlackHawk program \cite{2019EPJC...79..693A}, shows a consistency at the factor of 2 level in the H.E.S.S. energy range.

The theoretical 
% Probably average needed
average 
number of gamma rays detected from a PBH located at a distance $r_0$ and direction ($\mathrm{\alpha}_0$,$\mathrm{\delta}_0$) in the sky, during the last $\mathrm{\Delta}$t seconds of its life is given by:
%\begin{equation}
%N_{\mathrm{\gamma}}(r,\mathrm{\alpha},\mathrm{\delta},\mathrm%{\Delta}t) = \frac{1}{4\mathrm{\pi}r^{2}} %\int_{0}^{\mathrm{\Delta}t}dt \int_{0}^{\infty} %dE_{\mathrm{\gamma}} %\frac{d^{2}N}{dE_{\mathrm{\gamma}}dt}(E_{\mathrm{\gamma}},t) %A(E_{\mathrm{\gamma}},\mathrm{\alpha},\mathrm{\delta}),
%\end{equation}
\begin{equation}
N_{\mathrm{\gamma}}(r_0,\mathrm{\alpha}_0,\mathrm{\delta}_0,\mathrm{\Delta}t) =  
\int_{0}^{\infty} dE_{\mathrm{\gamma}} \frac{dN (\Delta t)}{dE_{\mathrm{\gamma}}} A(E_{\mathrm{\gamma}},\mathrm{\alpha}_0,\mathrm{\delta}_0)\,,
\label{eq:NCounts}
\end{equation}
where 
%d$^{2}$N/dE$_{\mathrm{\gamma}}$dt 
$N(\Delta t) = N(>E)$ is defined in equations \ref{eq:highE} and \ref{eq:lowEsimple}.  
%The photon spectrum emitted during a PBH evaporation becomes harder in the very last seconds of the PBH's life. 
The break energy $Q$ increases  
during PBH's life
%The signature of a PBH evaporation is thus a short, few seconds long, burst of high energy photons.
%The break energy $Q$ 
and is larger than 1 TeV in the last $\sim 18$ hours.
%of a PBH's life. 
However, equations
\ref{eq:Qdef} and \ref{eq:lowEsimple} show that the number
of gamma rays from PBH evaporation scales like $\sqrt{\Delta t}$ while the background increases linearly with $\Delta t.$ Because of this, in practice, the optimal time-window $\Delta t$ for the search with H.E.S.S. is in the range of a few to a few tens of seconds. The signature of a PBH evaporation is thus a burst of high-energy photons lasting a few seconds. 
 
The evaporation photon spectrum has to be folded with the H.E.S.S. effective area A(E$_{\mathrm{\gamma}}$,$\mathrm{\alpha}_0,\mathrm{\delta}_0$) to take into 
account the instrument's efficiency in collecting gamma rays of energy E$_{\mathrm{\gamma}}$ at equatorial coordinates $(\mathrm{\alpha_0},\mathrm{\delta_0})$ in the sky. The response of the H.E.S.S. instrument to gamma rays depends on the 
zenith angle and 
%offset angle of observation.  
the event's offset to the camera center.
% VIM : Remove this sentence, does not seems needed :
%The effective area $ A(E_{\mathrm{\gamma}},\mathrm{\alpha}_0,\mathrm{\delta})_0$ is an average over many runs with different zenith and offset angle. 
%
%It can be approximately factorized into an energy dependent term  and a spatially dependent  term by writing
%$ A(E_{\mathrm{\gamma}},\mathrm{\alpha},\mathrm{\delta})=A_{(0)}(E_{\mathrm{\gamma}}) A_{(1)}(\mathrm{\alpha},\mathrm{\delta}).$ 
%The factor $A_{(1)}(\mathrm{\alpha},\mathrm{\delta})$ corresponds basically to the normalised sky acceptance map to $\gamma$-rays, which is maximum at the center of the camera and drops toward the edges. It is directly estimated from the data.   
%The $A_{(0)}$ term is the effective area and is obtained from photon simulations with different zenith angles and offsets. 

The probability of detecting a burst of size\footnote{
% VIM : 15/11/2022 : Change the active to passive voice.
%In the following, we denote the number of gamma rays detected in a burst as the ``size" of the burst
In the following, the number of gamma rays detected in a burst is named the size of the burst
} b when observing a PBH with expected number of detected gamma rays N$_{\mathrm{\gamma}}$($r_0$,$\mathrm{\alpha}_0$,$\mathrm{\delta}_0$,$\mathrm{\Delta}$t) follows the Poisson statistics:
\begin{equation}
P(b,N_{\mathrm{\gamma}}) = e^{-N_{\mathrm{\gamma}}} \frac{N_{\mathrm{\gamma}}^{b}}{b!}\,.
\end{equation}
%VIM : 15/11/2022 : probability over space --> over all space
Integrating this probability over all space and assuming an isotropic distribution of PBH sources gives the theoretical number of expected bursts of size b to be detected in the data for an observation run $i$:
\begin{equation}
n^i_{sig}(b,\mathrm{\Delta}t,\mathrm{\dot{\rho}_{PBH}}) = \mathrm{\dot{\rho}_{PBH}}V^i_{\mbox{eff}}(b,\mathrm{\Delta}t)\,,
\label{eq:ngammath}
\end{equation} 
where $\dot{\rho}_\mathrm{PBH}$ is the local PBH evaporation burst rate per unit volume and the effective space-time volume of PBH detection is defined by 
\begin{equation}
V^i_{\mbox{eff}}(b,\mathrm{\Delta}t) =
 t_{i} \int d\mathrm{\Omega_{i}} \int_{0}^{\infty} dr r^{2} P_{i}(b,N_{\mathrm{\gamma}})\,,
\label{eq:effectivevolume}
\end{equation}
where the index i goes over each run of the H.E.S.S. data set, t$_i$ and $d\mathrm{\Omega_i}$ being the corresponding run live time and observation solid angle, respectively.\\

The effective volume can be written explicitly as
\begin{equation}
    V^i_{\mbox{eff}}(b,\mathrm{\Delta}t) =
    %T_{\mathrm{run}} \rho_{\mathrm{PBH}}
     t_{i} \mathrm{\Omega_{i}}
    \frac{{(r_0 \sqrt{N_0})}^3}{2} \frac{ \Gamma(b-3/2) }{\Gamma(b+1) }\,,
    \label{eq:veff}
\end{equation}
where $N_0$ is the expected number of photons from a PBH at $r_0$ and $\Gamma$ is Euler's gamma function. 

\section{Data analysis}\label{sec:analysis}

The signature of a PBH evaporation is thus a burst of  a small number ($\ge 2$) of gamma-like events which 
%The aim of this analysis is to estimate the excess of clusters of few gamma-like events (2 to 15) which 
arrive in coincidence in angular space and time  within a few seconds. This section describes the algorithm used to find the photon clusters in the H.E.S.S. data set, the estimation of the background rate and the statistical methods used to put 
constraints
%upper limits 
on the PBH evaporation rate.

\subsection{Clustering algorithm}
The clustering method used in this %analysis 
paper
is 
%strongly 
based on the OPTICS (Ordering Points To Identify the Clustering Structure) algorithm \cite{OPTICS} 
%and use it's 
implemented in the scikit-learn library \cite{DBLP:journals/corr/abs-1201-0490}. 
%This algorithm is closely related to 
OPTICS is an improved version of
DBSCAN \cite{DBSCAN} which builds clusters step by step, grouping points that are closely packed together given a predefined maximum distance. %Unlike DBSCAN, the OPTICS algorithm finds core samples of high density and expands clusters from them.  
% VIM 15/11/2022 rephrase of the DBSCAN paragraph :
 Three parameters are required for DBSCAN: 
 $\epsilon_\theta$, the maximum angular distance on the sky between two points, $\epsilon_t$, the maximum duration between two events and MinPts, the minimum number of events to declare a cluster.
 In this analysis, MinPts has been set to 2, $\epsilon_t$ to $\Delta t$ and $\epsilon_\theta$ to $0.14^\circ,$ which is twice the integration radius used in the point-like analysis with ImPACT reconstruction.
%DBSCAN requires two parameters: $\epsilon,$ the maximum distance to consider, and MinPts, the minimal number of points required to form a cluster. 
% VIM : 15/11/2022 : Change the sentence to be passive
%In our case we use a minimum number of points MinPts = 2 and a maximum distance $\epsilon_\theta = 0.14^\circ,$ which is twice the integration radius used in point-like ImPACT analysis, for the space dimension and $\epsilon_t = \Delta t$ for the time dimension. 
%The selected maximum distance, $\epsilon_\theta = 0.14^\circ,$ which is twice the integration radius used in point-like ImPACT analysis, for the space dimension and $\epsilon_t = \Delta t$ for the time dimension. 
%MinPts has been set to 2.
%The maximum chosen distance, $\epsilon = 0.14^\circ,$ is twice the integration radius used in point-like ImPACT analysis, for the space dimension and $\epsilon_t = \Delta t$ for the time dimension. 
OPTICS is more flexible than DBSCAN and allows finding clusters in a variable density environment. 
% VIM : 20 Oct 2022 :", which is twice the typical PSF" --> ", which is twice the integration radius used in point-like ImPACT analysis".
It finds core samples of high density and expands clusters from them.  OPTICS uses a fourth parameter $\xi$ which defines the density steepness at the cluster boundary.
% This xi definition comes from OPTICs scikit-learn documentation.
%For our analysis,
% VIM : 15/11/2022 : Change for passive voice
%We set $\xi$ to the default value $\xi = 0.05.$
The parameter $\xi$ has been set to the default value $\xi = 0.05.$

%It also  considers a reachability parameter which is based on the minimum distance between two event in the cluster and reflects the remoteness of an event given a cluster density.

%Like DBSCAN, OPTICS requires two parameters: $\epsilon$, which describes the maximum distance to consider, and MinPts, describing the number of points required to form a cluster. 

%OPTICS algorithm allows to control the reachability steepness of the clusters using a third parameter $\xi$. For our analysis, we found that $\xi = 0.05$ gives the expected output.

Clusters found may be spatially larger than the point-like integration radius or last longer than the chosen $\Delta t$.
When this happens, photon events from the cluster which are distant from the cluster median position in angle by more than this radius are excluded.
The procedure is iterated until the cluster smallest enclosing circle reaches this radius, which makes it then comparable to the effective area derived by Monte-Carlo simulations (cf equation \ref{eq:NCounts}).
The same procedure is performed in the time dimension. 
% VIM : 20 Oct 2022 : Original paragraph : 
% Clusters found may be spatially larger than the H.E.S.S. PSF or last longer than the chosen $\Delta t$.
% When this happens, photon events from the cluster which are distant from the cluster median position in space by more than $0.07^\circ$ are excluded. The procedure is iterated until the cluster smallest enclosing circle reaches a radius  
% $< 0.07^\circ$. The same procedure is performed in the time dimension. 

%After this first pass,
%the same algorithm is run a second time on the photons which are not associated to any cluster.
After excluding photons from large clusters, it happens that some groups of photon events still follow our cluster definition. A second pass of this same algorithm is then applied to the remaining events. The whole cluster-finding procedure was validated by injecting simulated clusters of photons in the data. The cluster-finding efficiency is estimated to be  $\simeq 98$\% for 5-photon clusters.

\subsection{Background estimation}

\begin{figure}[!h]
\centering
\begin{subfigure}{.5\textwidth}
  \centering
  \includegraphics[clip,width=\linewidth]{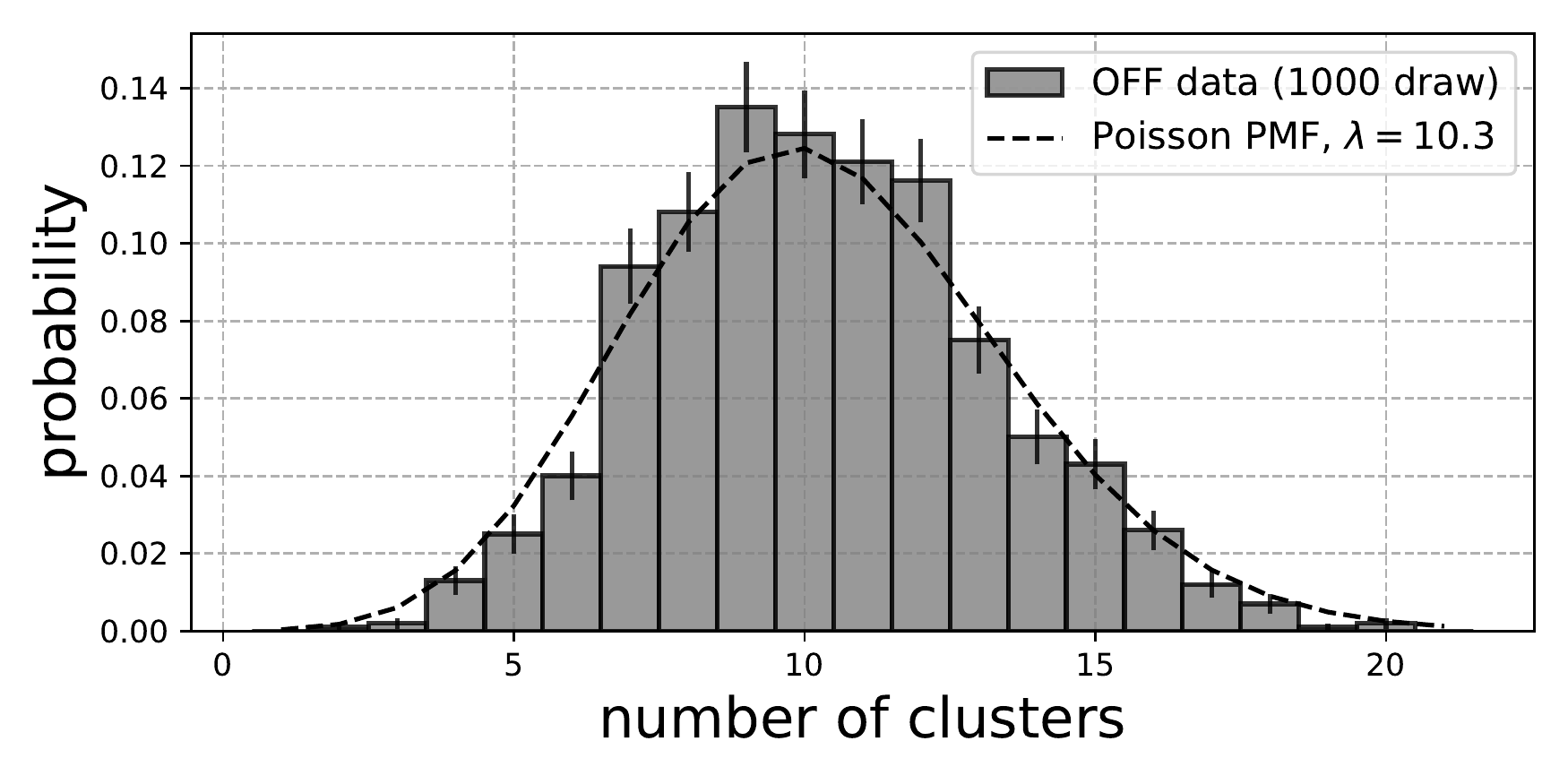}
  \caption{}
  \label{1gmes}
\end{subfigure}\hfill%
\begin{subfigure}{.5\textwidth}
  \centering
  \includegraphics[clip,width=\linewidth]{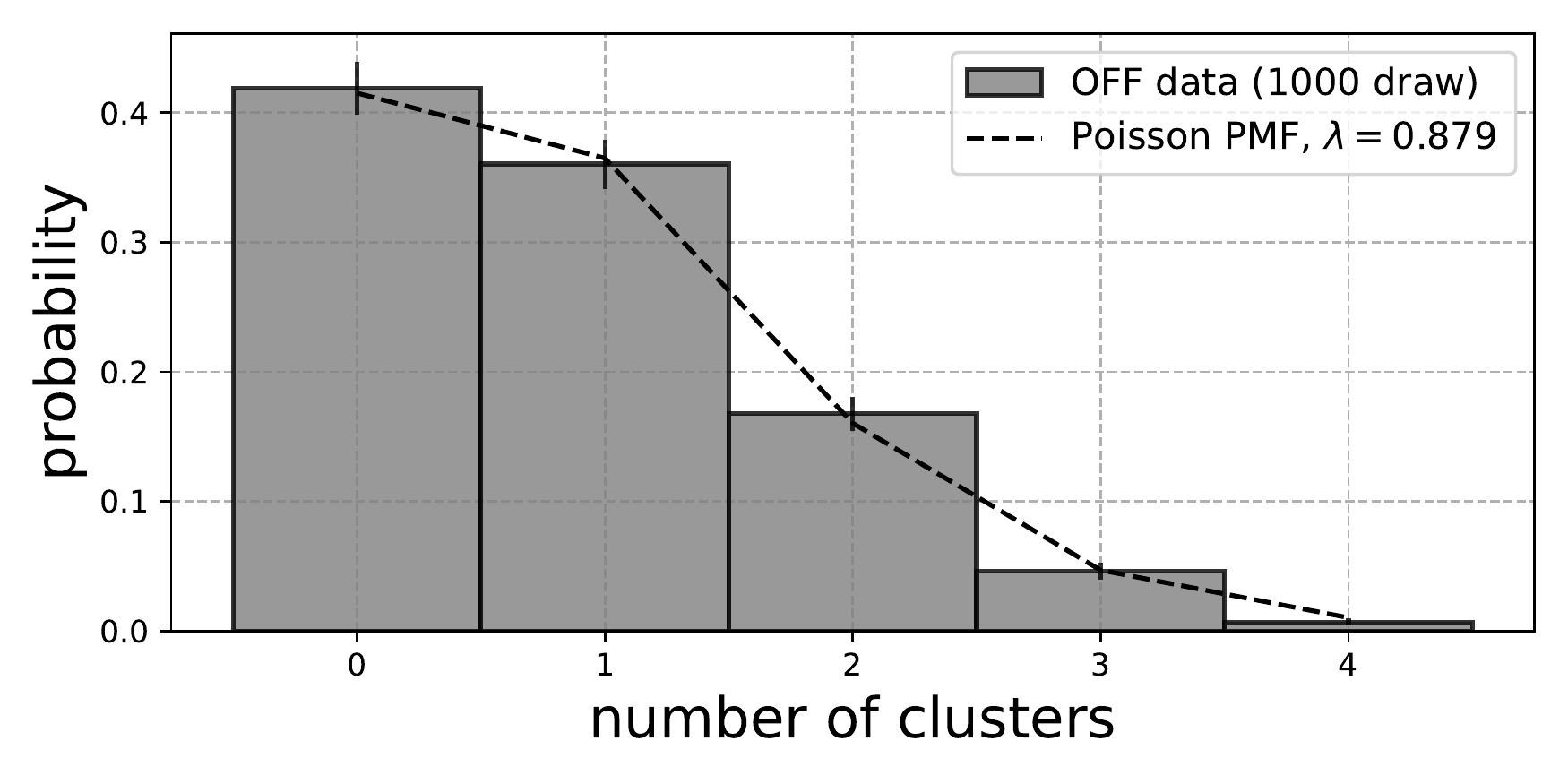}
  \caption{}
  \label{2gmes}
\end{subfigure}
\caption{Calculation of the statistical background for 60-second clusters  of %Distribution of the number of clusters found in 1000 random draws, for the cases of 
$b=4$ (a) and $b=5$ (b) photons. 
The photon list of a 28-minute H.E.S.S. run was time-scrambled 1000 times instead of the more usual 200 times. 
A Poisson law (dashed line) with an expected mean value $\lambda$
equal to the measured value in the OFF data is superimposed.
% VIM : 3 Nov 2022 : Original line : A Poisson law (dashed line) with a mean equal to the measured mean in the OFF data is superimposed.
}
\label{fig:poisson}
\end{figure}

Random fluctuations are expected to induce photon clusters by chance.
%It's expected to have an important number of clusters induced %by random fluctuations, 
It is therefore crucial to calculate the random cluster background with the best possible accuracy. The strategy followed in this paper is the direct estimation of the background from the data, by using the same photon list, but with randomized (``scrambled") times of arrival. The average value of the number of clusters distribution  %usually 
obtained by  time scrambling 200 times the photon list of each run (except for some runs, for which the number was increased to 1000) is taken as the background.
In the rest of the paper, the set of time-randomized photons will be referred to as OFF data. For a given cluster size, 
%(the number of photons in the cluster)
 the number of clusters found in the OFF data follows a Poisson distribution, as illustrated in Fig. \ref{fig:poisson}.

%The probability of  association of gamma-like events inside a cluster increases strongly with $\Delta T$.

\begin{figure}
  \centering
  \includegraphics[clip,width=.8\linewidth]{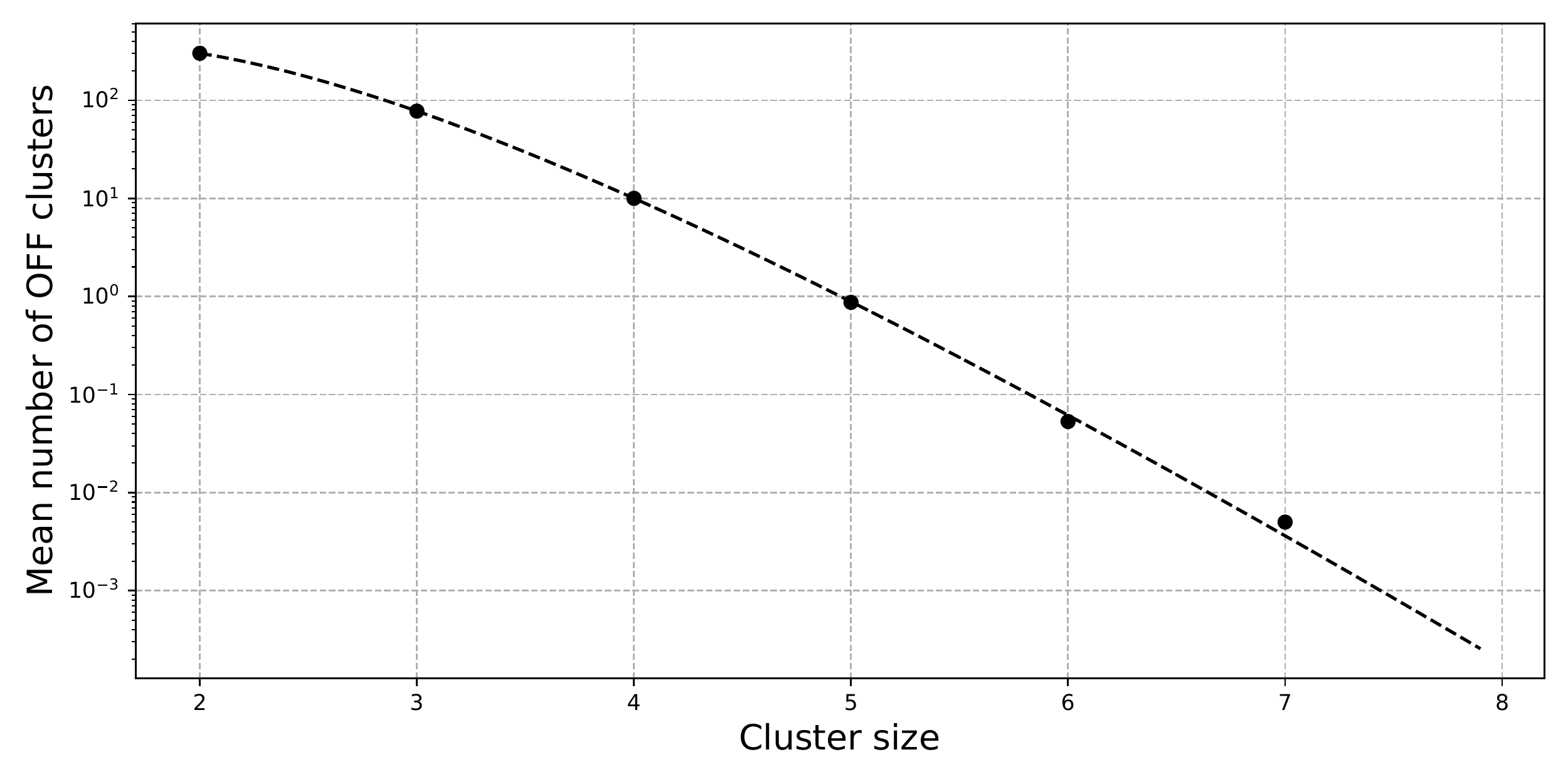}
  \caption{Distribution of the mean number of 60-second clusters found in 1000-time randomisation of a run (black dots). This distribution is fitted with a power-law with an exponential cut off (equation \ref{eq:ecpl}). The fitted distribution (black dashed line) is used to extrapolate the expected number of OFF clusters for large sizes.}
  \label{fig:extrap}
\end{figure}

The chance association of gamma-like events to form a cluster is a decreasing function of the photon cluster size. Therefore the OFF data are not sufficient to accurately predict the statistical background of large photon clusters.  To obtain the background value in that case, 
%the mean of the Poisson probability in such case,
the distribution of the mean number of photon clusters is fitted with a power-law distribution with an exponential cut off : % (equation \ref{eq:ecpl}).  

\begin{equation}
N(b) = A\  b^s e^{-t b}.
\label{eq:ecpl}
\end{equation}
The extrapolation of this equation is used  to obtain the mean of the Poisson probability distribution of cluster sizes 
%for each run.
for large cluster sizes.
Figure \ref{fig:extrap} shows an example of the distribution of cluster sizes for a H.E.S.S. run.

\subsection{Statistical methods}
%\subsection{Comparison of observed data to PBH signal expectations}

A hypothetical PBH signal is discovered or constrained by comparing the observed data to the
expected background (hypothesis $H_0$). The predicted flux depends on the unknown number density of bursting PBH, $\rho_\mathrm{PBH}.$ If an evaporation signal exists in the data, both the 
evaporation signal and the background will be observed (hypothesis $H_1$).
The PBH 
%evaporation 
density is estimated by maximizing a likelihood ratio with $\dot{\rho}_\mathrm{PBH}$ as the only free parameter, following the procedure of Feldman-Cousins \cite{FeldmanCousins}. The likelihood ratio is given by:

\begin{equation}
\frac{ { \cal L}_{H_1}} { { \cal L}_{H_0}} = 
\prod_{i} 
\frac{ { \cal P }  ( n^i_\mathrm{ON} | \lambda = n^i_\mathrm{OFF} +n^i_{sig}(b,\mathrm{\Delta}t, \dot{\rho}_\mathrm{PBH} )  ) }
{ { \cal P }  (n^i_\mathrm{ON} | \lambda = n^i_\mathrm{OFF} ) }\,,
\label{eq:Like_cousin}
\end{equation}
where ${ \cal P } $ is the Poisson probability,
$n^i_{sig}(b,\mathrm{\Delta}t,\dot{\rho}_\mathrm{PBH})$ is defined in equation \ref{eq:ngammath}, $n^i_\mathrm{ON}$ is the number of clusters found in the data and $n^i_\mathrm{OFF}$ is the corresponding mean number of clusters found in the OFF data.

The corresponding test statistic is given by:

\begin{equation}
TS =  -2 \ln \left(  \frac{ { \cal L} _{H_0}}{ { \cal L}_{H_1}} \right)  =  
-2 \times \sum_{i} \left[ n^i_{sig} + n^i_\mathrm{ON} \left( \ln(n^i_\mathrm{OFF}) - \ln ( n^i_\mathrm{OFF} + n^i_{sig}) \right) \right]\,.
\label{eq:TSM}
\end{equation}

%99\% CL upper limits are derived 

Another model-independent way to compare the data to the expected background is to look for a cluster excess of a given cluster size. 
For each time scale, $\Delta t$, and cluster size, $b$, the model-independent maximum-likelihood ratio test is given by

\begin{equation}
\frac{ { \cal L}_{H_1}} { { \cal L}_{H_0}} (b)= 
\prod_{i} 
\frac{ { \cal P }  ( n^i_\mathrm{ON}(b) | \lambda = n^i_\mathrm{OFF}(b) + n )}
{ { \cal P }  (n^i_\mathrm{ON}(b) | \lambda = n^i_\mathrm{OFF}(b) ) }\,,
\label{eq:Like_cousin_mi}
\end{equation}
where $n$ is the excess of clusters. Maximisation on the free parameter $n$ gives a model-independent estimation of the measured cluster excess for each cluster size, $b$.

\section{Results and discussion}

\subsection{Cluster distributions}\label{sec:results}
\begin{figure}[!h]
\centering
%\begin{subfigure}{.5\textwidth}
%  \centering
%  \includegraphics[clip,width=\linewidth]{./fig%/excessvssizedt30-v3.pdf}
%  \caption{}
%  \label{fig:excess30}
%\end{subfigure}\hfill%
%\begin{subfigure}{.5\textwidth}
%  \centering
  \includegraphics[clip,width=\linewidth]{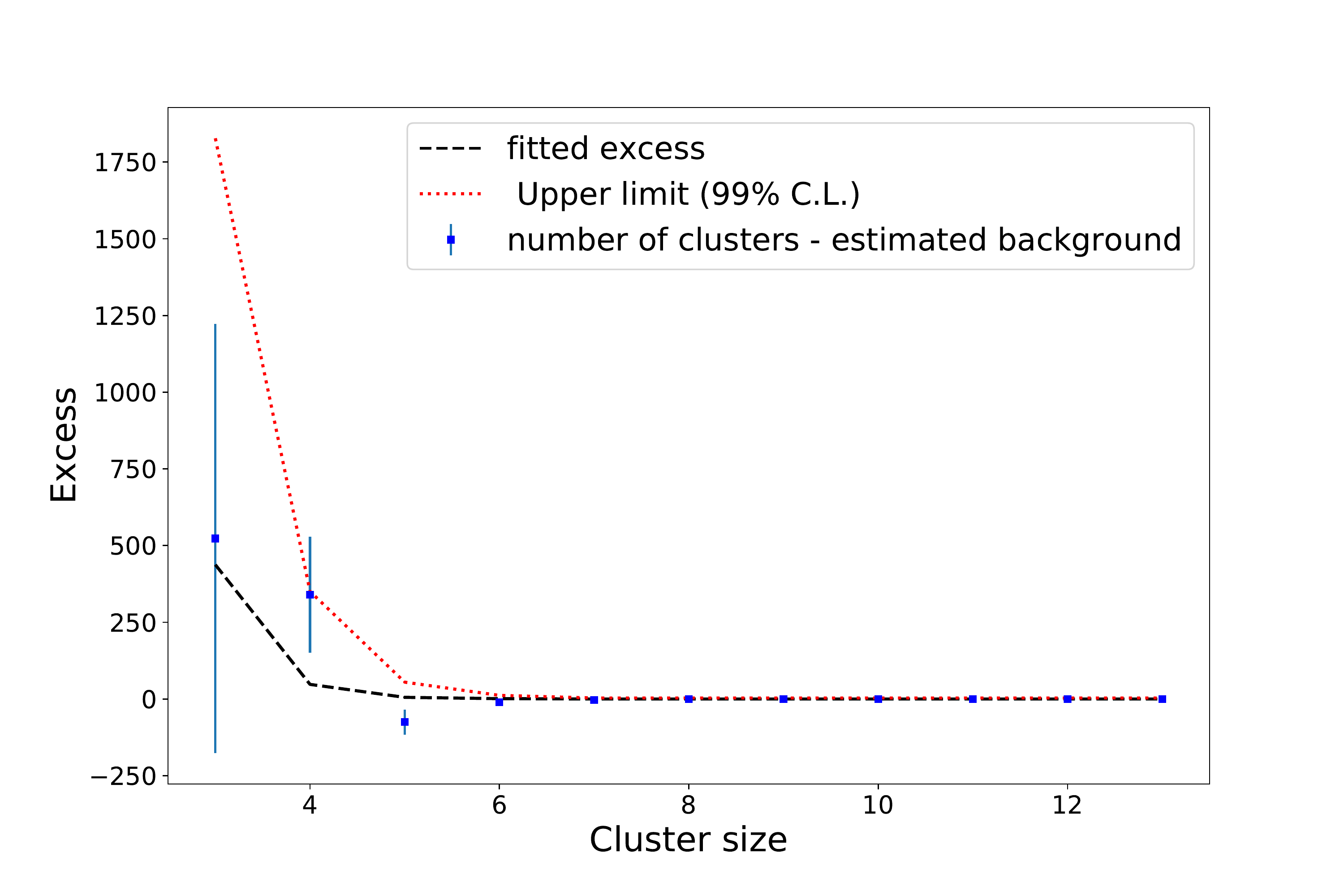}
%  \caption{}
%  \label{fig:excess60}
%\end{subfigure}
\caption{Cluster excess ($n_{\mathrm{ON}}-n_{\mathrm{OFF}}$), fitted excess (equation \ref{eq:Like_cousin})  and 99\% C.L. upper limit on the cluster excess (equation \ref{eq:Like_cousin_mi}) as function of cluster size
%(a): $\Delta t =30$ s, (b): 
for $\Delta t = 60$ s. 
%VIM 20 Dec 2022 : Attempt to remove ambiguity and clarify (to be improved)
The standard deviation of the  $n_{\mathrm{OFF}}$ distribution is shown as error bars.
%The square root of the average estimated of $N_{\mathrm{OFF}}$ is shown as error bar to illustrate typical background fluctuations.
}
\label{fig:excess60}
%\label{fig:excess}
\end{figure}

As discussed in section \ref{sec:model}, the size of the timescale $\Delta t$ is constrained by the statistical background. 
Therefore, the analysis was performed for 4 values of the timescale $\Delta t$, namely 10, 30, 60 and 120 seconds, using the 
%5335.1 
%4924 % Before VIM Analysis 
4816 % VIM Analysis
hours data set described in section \ref{sec:dataset}.
The lack of cluster statistics makes the 
background estimation difficult for timescales lower than 10 seconds. 
%Running the analysis described in the previous sections for timescales less than 10 seconds leads to a lack of statistics which makes it difficult to estimate the number of background clusters. 
The CPU load of the clustering algorithm increases with $\Delta t$ which led to limit the analysis to $\Delta t \leq 120$ seconds. 
%On the other hand, for timescales above 120 seconds, the CPU load of the clustering algorithm becomes too high to be reasonable.  
%The maximum value obtained for the test statistics (TS, eq. \ref{eq:TSM}) at any $\Delta t$ is $\sim 10^{-4}$. 
The difference between the number of clusters found in the ON and OFF data sets, as well as the fitted excess, obtained from the likelihood ration \ref{eq:Like_cousin}, and the 99\% upper limits on a cluster excess, derived from the likelihood ratio \ref{eq:Like_cousin_mi}, 
are  shown in Figure
%\ref{fig:excess30} and 
\ref{fig:excess60} for 
%$\Delta t=~30$~s and 
$\Delta t= 60$~s.

\begin{figure}[h!]
  \centering
 \includegraphics[clip,width=\textwidth]{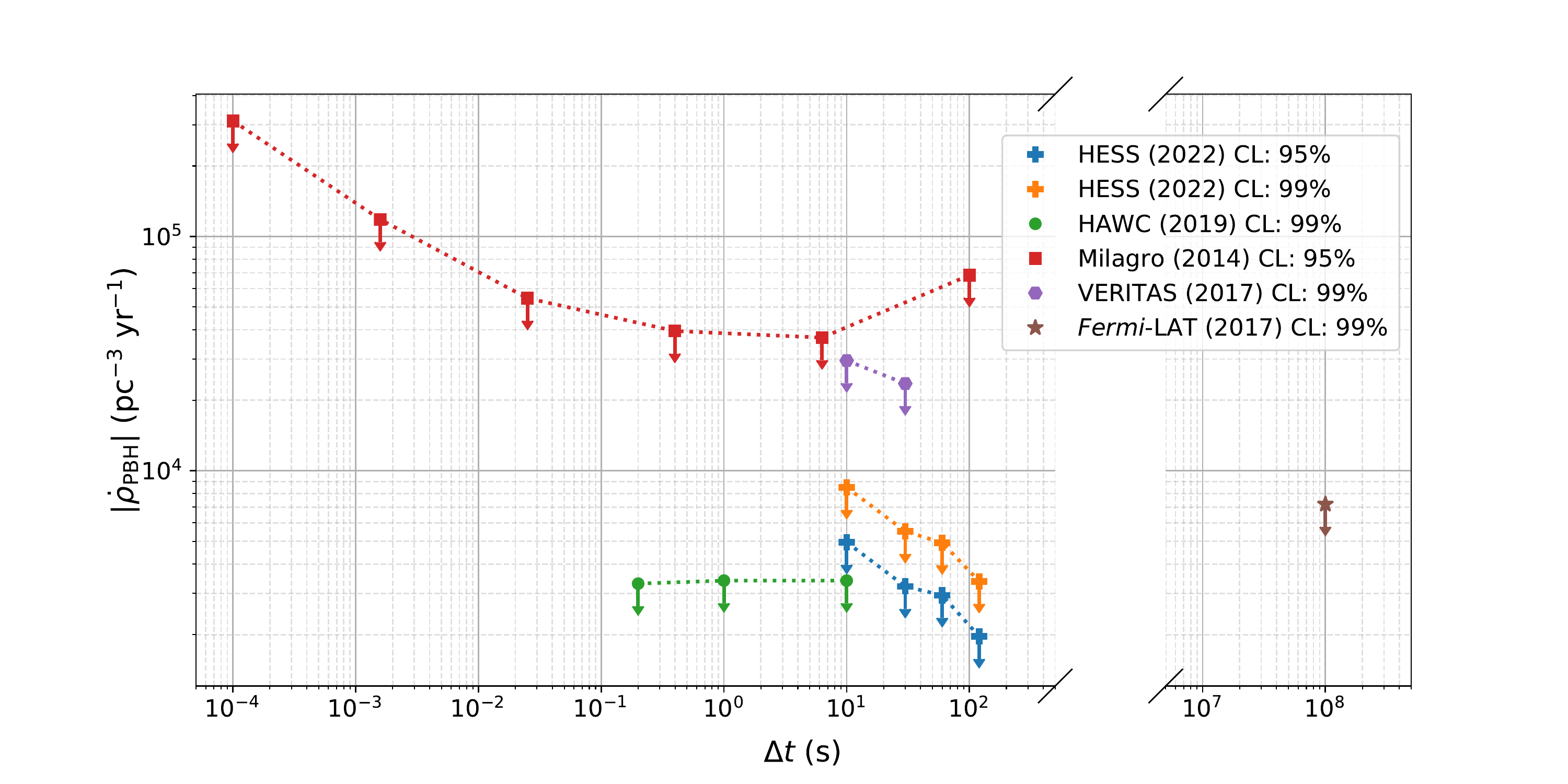}
  \caption{Upper limits on the PBH evaporation burst rate $\dot{\rho}_\mathrm{PBH}$ for burst time scales of 10, 30, 60 and 120 seconds measured by H.E.S.S. Upper limits from HAWC \cite{2020JCAP...04..026A}, VERITAS \cite{2017ICRC...35..691A},  Milagro \cite{abdo_milagro_2015} and  Fermi-LAT \cite{2018ApJ...857...49A} are also shown. 
  %The upper limits are inversely proportional to the effective volume (equation \ref{eq:veff}). A factor of 2 uncertainty in the predicted photon flux from a PBH evaporation translates into a factor 3 uncertainty in the limits.
  }
  \label{fig:results}
\end{figure}

No significant signal was found in the data. Upper limits on the PBH evaporation burst rate $\dot{\rho}_\mathrm{PBH}$ with  confidence levels (CL) of 95\% and 99\% were derived by finding the $\dot{\rho}_\mathrm{PBH}$ for
which TS = 3.84 and 6.63 respectively. These upper limits are shown in Fig. \ref{fig:results} and compared to the limits obtained by HAWC \cite{2020JCAP...04..026A}, VERITAS \cite{2017ICRC...35..691A},  Milagro \cite{abdo_milagro_2015} and  Fermi-LAT \cite{2018ApJ...857...49A}. Note that the limits depend on the predicted photon flux. The predictions used by H.E.S.S. are in fair agreement with those used by the other Cherenkov arrays and with those of the BlackHawk program \cite{2019EPJC...79..693A}. The Fermi-LAT result, which is derived from lower energy photons, uses a $\simeq 40$\% smaller photon flux. 
%FB : next 2 sentetnces were moved from fig caption:
The upper limits are inversely proportional to the effective volume (equation \ref{eq:veff}). A factor of 2 uncertainty in the predicted photon flux from a PBH evaporation translates into a factor 3 uncertainty in the limits. 

\subsection{Discussion}\label{sec:discussion}

The time evolution of PBH masses is given by
\begin{equation}
\frac{dM}{dt} = -\frac{\alpha(M)}{M^2},
\label{eq:evaporation}
\end{equation}
where $\alpha(M)$ counts the degrees of freedom of the particles contributing to the energy loss as a function of the black-hole mass \cite{1991Natur.353..807H}.
The low-mass limit of the function $\alpha(M)$ strongly depends 
on the assumed particle physics model. 
%It's important to keep in mind that 
The upper limits on $\dot{\rho}_\mathrm{PBH}$ derived in section \ref{sec:results} 
%strongly depend on the assumed particle physics model used to model the PBH signal (section \ref{sec:model}).
were obtained with the standard model of
particle physics (section \ref{sec:model}).
Because of equation \ref{eq:evaporation},
the PBHs evaporating at the present moment
were initially in a very narrow range of masses if they
formed at about the same time.
At the cost of additional hypotheses on the PBH mass distribution, it is possible to use our measurements to constrain the initial contribution of these PBHs to the dark matter (DM). A recent review of PBH as dark matter is given in \cite{2020ARNPS..70..355C}. Popular models for the initial mass distribution of PBHs are the log-normal and the power-law distributions \cite{2018JCAP...01..004B}. 
The latter is the simplest model for the initial mass distribution of PBHs and describes their production in the early universe by scale-invariant Gaussian density perturbations. 
%Several models for the initial mass distribution of PBH exists. The simplest is produced by scale-invariant Gaussian density perturbation, 
It is given by \cite{1991Natur.353..807H}:

\begin{equation}
\frac{d\mathcal{N}_\mathrm{PBH}}{dM_i} = \frac{\rho_0}{M_*} \left( \frac{M_i}{M_*} \right)^{-\beta}\,,
\label{eq:mass_dist}
\end{equation}
where $M_i$ is the initial mass of PBHs, $M_* \simeq (0.5-1)\times 10^{15} \mbox{g}$ 
is the initial mass of a PBH that formed in the early universe and is, at the present time, at the final stage of its evaporation
and { $\beta$ is the index of the initial PBH mass function, which is constrained to be larger than 2 and has a value of 2.5 in the radiation-dominated era \cite{2018JCAP...01..004B}.}
In equation \ref{eq:mass_dist}, 
$\mathcal{N}_\mathrm{PBH}$ is 
%$\rho_\mathrm{PBH}$ is 
%as before 
the number density of PBHs.

The normalization factor $\rho_0$ is given by 
%$\int_{M_*}^{\infty} \frac{d\rho_\mathrm{PBH}}{dM_i} M_i dM_i= \Omega_\mathrm{PBH} \rho_c.$

\begin{equation}
\rho_0 = (\beta - 2 ) \frac{\Omega_\mathrm{PBH} \rho_c}{M_*}\,,
\label{eq:rho_zero}
\end{equation}
where $\Omega_\mathrm{PBH} $ is the fraction of the critical density $\rho_c$ in PBHs with mass larger than $M_*$.
Equation 12 from \cite{1991Natur.353..807H} gives the current local rate of vanishing PBHs as:
\begin{equation}
|\dot{\rho}_\mathrm{PBH}| \simeq \frac{\alpha(M_*)}{M_*^3}\eta \rho_0\,, 
\label{eq:halzen_2}
\end{equation}
where 
%$\alpha(M)$ counts the degrees of freedom of the particles contributing to the energy loss as a function of the black-hole mass \cite{1991Natur.353..807H} and 
$\eta,$ the clustering factor, is the ratio between the local and global dark matter densities. 
%assuming an isothermal Milky Way halo.

\begin{figure}[h!]
  \centering

  \includegraphics[clip,width=.9\linewidth]{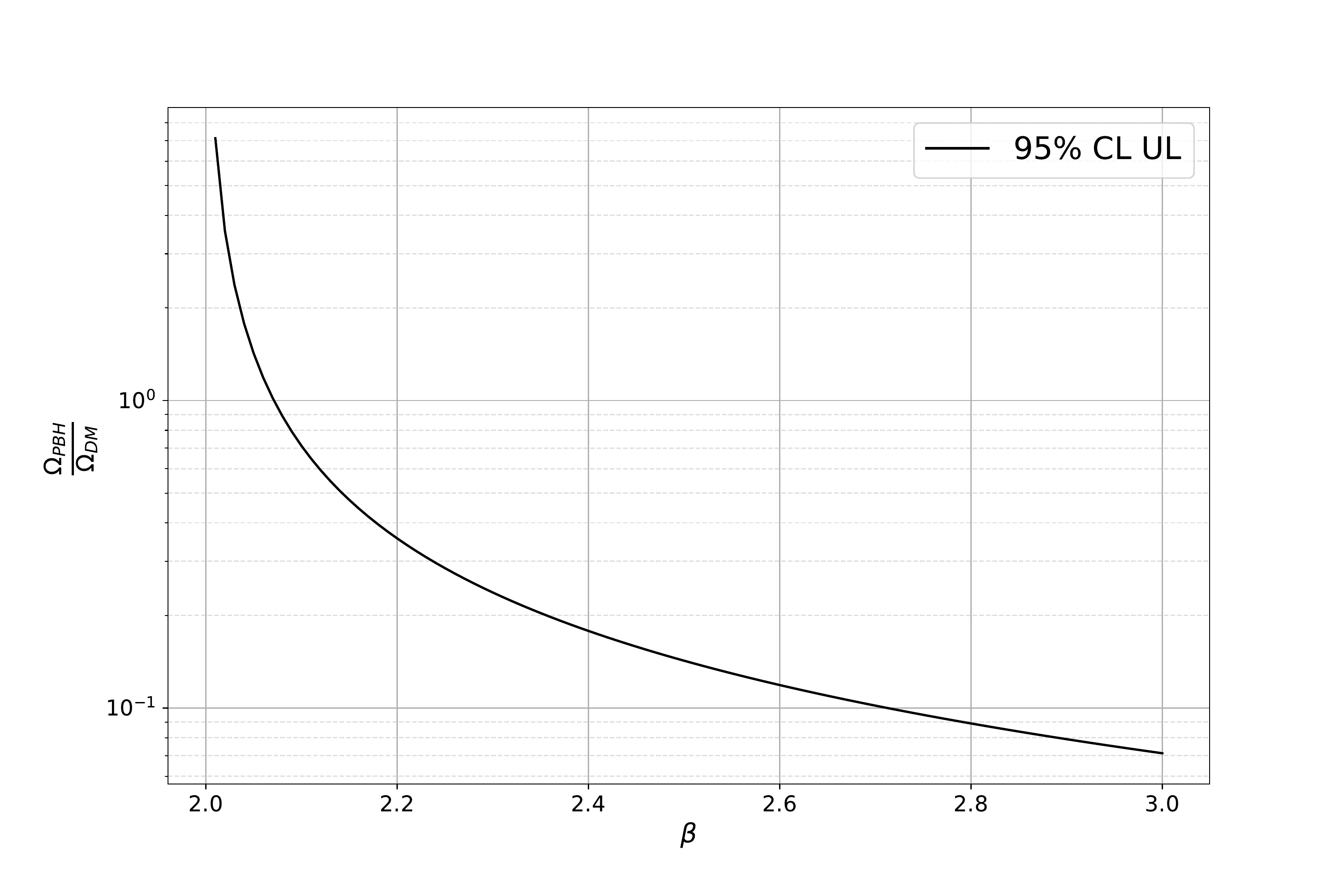}
  \caption{Upper limits on the initial PBH fraction of the dark matter density as a function of the PBH mass distribution index $\beta$.}
  \label{fig:ul_omega}
\end{figure}

Upper limits on the initial 
%{ and current} 
PBH mass fraction $\Omega_{\rm PBH}$ as a function of the index $\beta$ can be obtained from equations \ref{eq:rho_zero} and \ref{eq:halzen_2} taking, for $M_{\star}=10^{15}\ \mathrm{g},$ $\eta$ to be its 95 \% CL limit value $\eta > 1.6 \times 10^4$ \cite{bovy_local_2012} and 
%for $\alpha(M_*)$ the constraint 
$\alpha(M_*) > 10^{17}$ kg$^3$ s$^{-1}$ \cite{1991Natur.353..807H}.
%Using $M_{\star}=10^{15}\ \mathrm{g}$ and taking for $\eta$ its 95 \% CL limit values $\eta > 1.6 \times 10^4$ \cite{bovy_local_2012} and for $\alpha(M_*)$ the constrains $\alpha(M_*) > 10^{17}$ kg$^3$ s$^{-1}$ \cite{1991Natur.353..807H}, 
%our results allow us to derive 
%upper limits on the initial 
%{ and current} 
%PBH mass fractions as a function of the index $\beta$ can be obtained from equations \ref{eq:rho_zero} and \ref{eq:halzen_2}.  
Results are shown in Figure \ref{fig:ul_omega}. In the scenario of PBH formation from density fluctuations, 
the { initial} mass fraction in PBHs  { with mass larger than $M_*$} is constrained to be a subdominant 
fraction of {the dark matter} for most values of the $\beta$ 
exponent.
%it would require very fine tuning for these object to be responsible for the major part of the invisible mass.

%\section*{acknowledgments}

\begin{figure}[h!]
  \centering
  \includegraphics[clip,width=.9\linewidth]{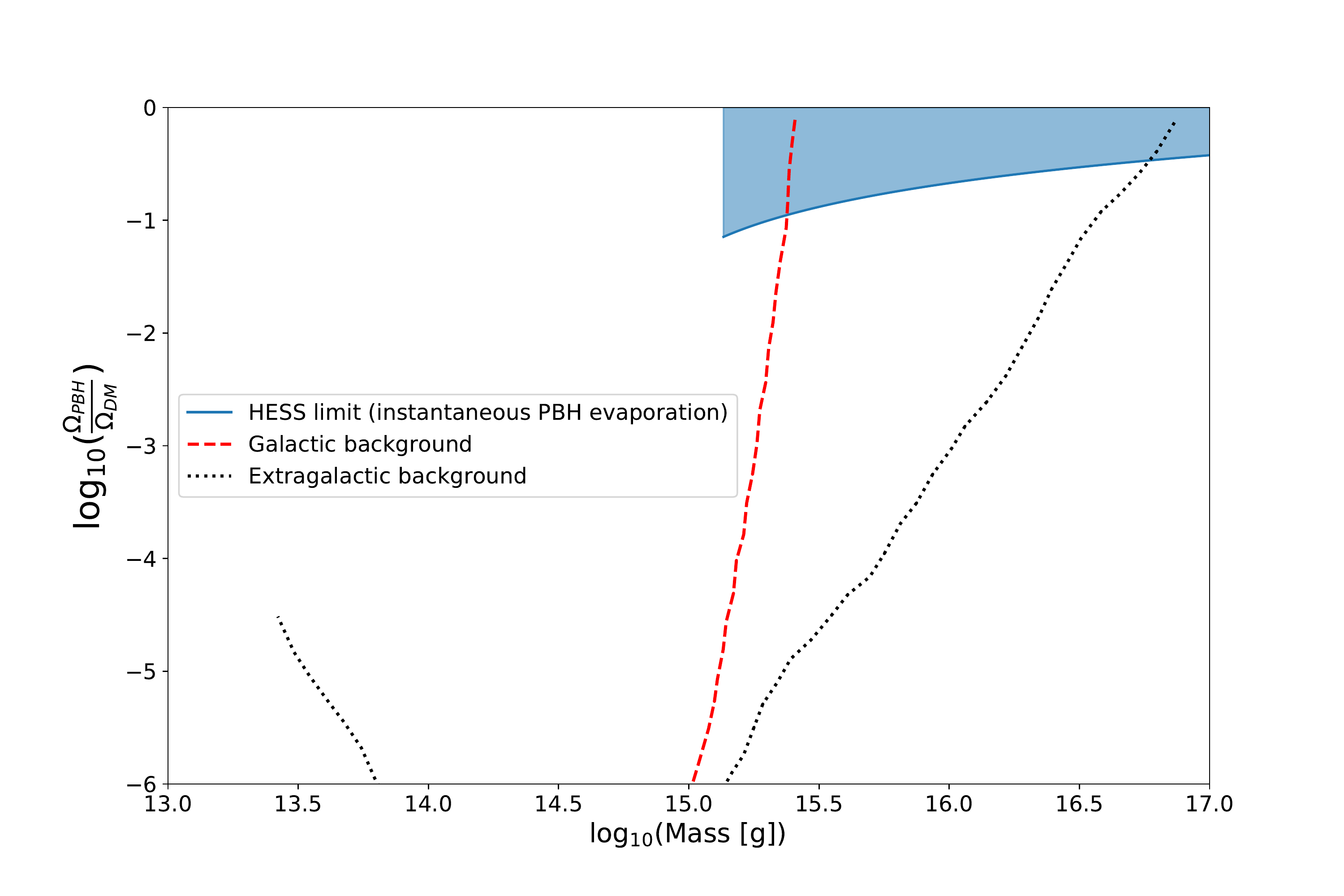}
  \caption{Upper limits on the present mass fraction of PBH as a function of the effective PBH mass. 
  The initial mass distribution index  $\beta$ is varied between 2 and 3, giving the blue exclusion region. 
  The limit is compared to monochromatic gamma-ray limits compiled in \cite{2021RPPh...84k6902C}.
  %The solid line is the H.E.S.S. limit, the dashed line is the extra-galactic background limit taken from Fig. 1 of \cite{2017PhRvD..96b3514C} and the dotted line shows the femtolensing limit \cite{2012PhRvD..86d3001B}. The validity of the femtolensing limit is debated due to the effects of the final size of the GRB sources  %\cite{2018JCAP...12..005K}.
  }
  \label{fig:ul_omegaM}
\end{figure}

Alternatively, fixing $\beta$ gives a constraint on the present
density of low-mass PBHs.
Figure \ref{fig:ul_omegaM} shows the 95\% confidence limits on the present density of PBHs derived from the limit on the present rate of evaporation bursts. 
$\beta$ was varied between 2 and 3. For each value of $\beta,$ an effective monochromatic PBH mass was calculated according to the prescription of \cite{2017PhRvD..96b3514C}.
The burst limit is compared to limits obtained from the Galactic and extra-galactic gamma-ray backgrounds
taken from \cite{2021RPPh...84k6902C}.
%taken from Fig. 1 of \cite{2017PhRvD..96b3514C} and the lack of femtolensed GRBs \cite{2012PhRvD..86d3001B}. 
%In Figure
%\ref{fig:ul_omegaM}, the initial mass distribution index $\beta = 3$ is assumed. 
The H.E.S.S. limit is much less constraining than 
%the limits 
those derived from the gamma-ray background measurements. But the H.E.S.S. measurement is local, sensitive to PBHs located within 0.1 pc from the Sun,  and is sensitive only to the initial value  $\mathrm{M}_{\star}$ of presently evaporating PBHs. In contrast, 
gamma-ray background 
limits on $\Omega_{\rm PBH}$
%measurements 
are sensitive to PBHs in much larger volumes, but also to details of the low-mass tail of the initial and evolved mass distribution of PBHs.

%\subsection{Conclusion}\label{sec:conclusion}
\section{Conclusions}\label{sec:conclusion}
The theory of Hawking radiation predicts that PBHs at their final stage of evaporation should produce bursts of high-energy particles.
%4924 % Before VIM Analysis 
%4816 %VIM Analysis
More than 4800
hours of H.E.S.S. observations have been used to search for short time-scale (10 s to 120 s) clusters of photons corresponding to the expected evaporation burst signal. The number of clusters found is fully compatible with statistical fluctuations with no significant signal. The most constraining 95\% CL upper limit on the PBH burst rate was found to be $|\dot{\rho}_\mathrm{PBH}| <
%1965
2000
$ pc$^{-3}$ yr$^{-1}$. This limit is 
comparable to those obtained by HAWC
\cite{2020JCAP...04..026A}
and Fermi-LAT \cite{2018ApJ...857...49A}, taking into account the different signal prediction. 
%% VIM : 30 November 2022 : Original Line
Finally, the limit on the PBH evaporation rate has been translated into a constraint on the initial fraction of dark matter in PBHs of order $0.1\ \Omega_{DM}$ for the hypothesis of a PBH initial mass power-law distribution.
In the near future, the upcoming VHE gamma ray observatory CTA will be sensitive to a much larger fraction of the sky and should be able to lower the limit on the PBH burst rate obtained in the present paper by an order of magnitude if no bursts are detected
\cite{2021arXiv211101198D}.
%\textbf{VIM Note: The DM sentence needs to be expanded because it does not conclude/summarize anything}

\acknowledgments

{\small The support of the Namibian authorities and of the University of
Namibia in facilitating the construction and operation of H.E.S.S.
is gratefully acknowledged, as is the support by the German
Ministry for Education and Research (BMBF), the Max Planck Society,
the German Research Foundation (DFG), the Helmholtz Association,
the Alexander von Humboldt Foundation, the French Ministry of
Higher Education, Research and Innovation, the Centre National de
la Recherche Scientifique (CNRS/IN2P3 and CNRS/INSU), the
Commissariat \`a l’\'energie atomique et aux \'energies alternatives
(CEA), the U.K. Science and Technology Facilities Council (STFC),
the Irish Research Council (IRC) and the Science Foundation Ireland
(SFI), the Knut and Alice Wallenberg Foundation, the Polish
Ministry of Education and Science, agreement no. 2021/WK/06, the
South African Department of Science and Technology and National
Research Foundation, the University of Namibia, the National
Commission on Research, Science \& Technology of Namibia (NCRST),
the Austrian Federal Ministry of Education, Science and Research
and the Austrian Science Fund (FWF), the Australian Research
Council (ARC), the Japan Society for the Promotion of Science, the
University of Amsterdam and the Science Committee of Armenia grant
21AG-1C085. We appreciate the excellent work of the technical
support staff in Berlin, Zeuthen, Heidelberg, Palaiseau, Paris,
Saclay, T\"ubingen and in Namibia in the construction and operation
of the equipment. This work benefited from services provided by the
H.E.S.S. Virtual Organisation, supported by the national resource
providers of the EGI Federation.}

\bibliography{pbh}{}
\bibliographystyle{JHEP}

\clearpage
%VIM : Force the appendice to start on a new page.
%-------------------------------------------------------------------------------------------------------
%-------------------------------------------------------------------------------------------------------
%\appendix

%\section{Results tables}
%\begin{center}
%\begin{table}[h!]
%\begin{tabular}{||c c c c c c||}
%\hline
%Cluster Size & $N_\mathrm{ON}$  & $N_\mathrm{OFF}$ &  %excess & fitted exces & 99\% CL UL \\
%\hline \hline
%2 & 883782.0 & 882853.4900 & 928.5100 & 1177.3899 & %3172.8952  \\
%\hline
%3 & 28529.0 & 28635.7350 & -106.7350 & -63.5653 & %208.7587  \\
%\hline
%4 & 416.0 & 464.6500 & -48.6500 & 2.9030 & 36.7293  \\
%\hline
%5 & 2.0 & 15.0490 & -13.0490 & -0.0440 & 3.9697  \\
%\hline
%6 & 0.0 & 1.6309 & -1.6309 & -0.0000 & 3.6450  \\
%\hline
%7 & 0.0 & 0.3640 & -0.3640 & -0.0000 & 3.6450  \\
%%\hline
%8 & 0.0 & 0.1060 & -0.1060 & -0.0000 & 3.6450  \\
%%\hline
%9 & 0.0 & 0.0382 & -0.0382 & -0.0000 & 3.6450  \\
%\hline
%10 & 0.0 & 0.0174 & -0.0174 & -0.0000 & 3.6450  \\
%\hline
%11 & 0.0 & 0.0103 & -0.0103 & -0.0000 & 3.6450  \\
%\hline
%12 & 0.0 & 0.0076 & -0.0076 & -0.0000 & 3.6450  \\
%\hline
%\end{tabular}
%\caption{ Result table for 10 seconds time scale.}
%\label{table:10s}
%\end{table}
%\end{center}

%------------------------------------------------------------------------------------------------------   1231.3450  

\end{document}